\documentclass[12pt,epsf]{article}
\usepackage{verbatim}

\usepackage{dsfont}

\usepackage[english]{babel}

\usepackage{amssymb,amsmath}
\usepackage{graphicx, xcolor, varwidth}
\usepackage{setspace}
\usepackage[permil]{overpic}
\usepackage{cite}

\colorlet{darkblue}{blue!70!black}
\newcommand{\arXiv}[1]{\href{http://www.arXiv.org/abs/#1}{#1}}
\usepackage[colorlinks=true,urlcolor=darkblue,linktocpage=true,linkcolor=darkblue,citecolor=darkblue]{hyperref}

\numberwithin{equation}{section}

\newcommand{\be}{\begin{equation}}
\newcommand{\ee}{\end{equation}}
\newcommand{\bea}{\begin{eqnarray}}
\newcommand{\eea}{\end{eqnarray}}
\newcommand{\bear}{\begin{eqnarray}}
\newcommand{\eear}{\end{eqnarray}}
\newcommand{\beas}{\begin{eqnarray*}}

\newcommand{\eeas}{\end{eqnarray*}}
\newcommand{\ba}{\begin{array}}
\newcommand{\ea}{\end{array}}

\newcommand{\tr}{\operatorname{tr}}
\newcommand{\pd}[2][1]{\ifnum#1=1 \frac{\partial}{\partial {#2}} \else
  \frac{\partial^#1}{\partial {#2}^{#1}}\fi}
\newcommand{\dpd}[2][1]{\ifnum#1=1 \dfrac{\partial}{\partial {#2}} \else
  \frac{\partial^#1}{\partial {#2}^{#1}}\fi}
\newcommand{\td}[2][1]{\ifnum#1=1 \frac{d}{d{#2}} \else
  \frac{d^#1}{d{#2}^{#1}}\fi}

\renewcommand{\(}{\left(}
\renewcommand{\)}{\right)}

\newcommand{\nbox}{{\,\lower0.9pt\vbox{\hrule \hbox{\vrule height 0.2 cm \hskip 0.19 cm \vrule height 0.2 cm}\hrule}\,}}
\newcommand{\Tr}{\ {\rm Tr}\ }

\newcommand{\eps}{\epsilon}

\newcommand{\ie}{{\it i.e.,}\ }

\newcommand{\bL}{\mathbf{L}}

\renewcommand{\bL}{\overline{L}}

\newcommand{\bh}{\bar{h}}

\newcommand{\bz}{\bar{z}}
\newcommand{\tsigma}{\tilde{\sigma}}
\newcommand{\bH}{\overline{H}}
\newcommand{\bw}{\bar{w}}
\newcommand{\bF}{\mathcal{F}}
\newcommand{\bO}{\mathcal{O}}
\newcommand{\bT}{\overline{T}}
\newcommand{\bbeta}{\bar{\beta}}
\newcommand{\bdelta}{\bar{\delta}}
\newcommand{\balpha}{\bar{\alpha}}



\begin{document}
\begin{spacing}{1.3}
\begin{titlepage}

\begin{center}
{\Large \bf Holographic Entanglement Entropy from 2d CFT: 
\\ 
\vspace{.3cm}
Heavy States and Local Quenches}

\vspace*{6mm}

Curtis T.~Asplund$^*$, Alice Bernamonti$^\ddag$,  Federico Galli$^\ddag$, and Thomas Hartman$^\S$
\vspace*{6mm}

\textit{$^*$ Department of Physics, Columbia University, New York, New York 10027, USA}\\ 
\textit{$^\ddag$ Instituut voor Theoretische Fysica, KU Leuven, Celestijnenlaan 200D,\\ B-3001 Leuven, Belgium}\\
\textit{$^\S$ Department of Physics, Cornell University, Ithaca, New York 14853, USA}\\ 

\vspace{6mm}

{\tt ca2621@columbia.edu, alice@itf.fys.kuleuven.be, federico@itf.fys.kuleuven.be, hartman@cornell.edu}

\vspace*{6mm}
\end{center}
\begin{abstract}

We consider the entanglement entropy in 2d conformal field theory in a class of excited states produced by the insertion of a heavy local operator. These include both high-energy eigenstates of the Hamiltonian and time-dependent local quenches. We compute the universal contribution from the stress tensor to the single interval Renyi entropies and entanglement entropy, and conjecture that this dominates the answer in theories with a large central charge and a sparse spectrum of low-dimension operators. The resulting entanglement entropies agree precisely with holographic calculations in three-dimensional gravity. High-energy eigenstates are dual to microstates of the BTZ black hole, so the corresponding holographic calculation is a geodesic length in the black hole geometry;  agreement between these two answers demonstrates that entanglement entropy thermalizes in individual microstates of holographic CFTs. 
For local quenches, the dual geometry is a highly boosted black hole or conical defect.  On the CFT side, the rise in  entanglement entropy after a quench is directly related to the monodromy of a Virasoro conformal block.

\end{abstract}

\end{titlepage}
\end{spacing}

\vskip 1cm

\tableofcontents

\begin{spacing}{1.3}

\section{Introduction}

Certain conformal field theories (CFTs) are dual to quantum gravity in asymptotically anti-de Sitter (AdS) spacetime.  These CFTs must have a large $N$ limit for the bulk to be weakly coupled, and a sufficiently sparse spectrum of low-dimension operators to ensure that the bulk effective field theory has a limited number of fields.  It is suspected that any CFT with these two properties has a holographic dual, and indeed, these two criteria alone are enough to begin building the bulk effective field theory perturbatively in $1/N$ about the vacuum state directly from CFT \cite{Heemskerk:2009pn, Heemskerk:2010ty, Fitzpatrick:2010zm, Fitzpatrick:2013twa}. However, to derive other universal features of AdS quantum gravity, such as thermal free energy, transport coefficients, and entanglement entropy, new methods are needed to reorganize CFT calculations at high energy density. Holography suggests that at least within this class of theories, these methods should exist and should dramatically simplify the structure of excited states in much the same way that the Wilsonian renormalization group simplifies the ground state of a local Hamiltonian (e.g., \cite{White:1992zz}).

In two-dimensional CFT, infinite-dimensional conformal symmetry simplifies the problem of studying these non-trivial classical states carrying $O(N)$ energy.  It has been used, for example, to set quantitative bounds on the required sparsity of the spectrum to ensure that the bulk dual has black holes with semiclassical thermodynamics \cite{hks}, and to derive from CFT the entanglement entropy in vacuum \cite{Headrick:2010zt,Faulkner:2013yia,tom,Barrella:2013wja,Chen:2013kpa,Perlmutter:2013paa}.

Entanglement entropy is also useful for studying quantum field theories in excited states, 
in both static and dynamical situations. 
The geometric entanglement entropy
in the CFT is related to the bulk through the holographic entanglement entropy 
proposal \cite{Ryu:2006bv,Hubeny:2007xt,Lewkowycz:2013nqa}. On the gravity side, this is well established in non-dynamical 
situations \cite{Headrick:2007km, Hayden:2011ag, Lewkowycz:2013nqa}, and evidence is accumulating in its favor in dynamical spacetimes, e.g.  \cite{Callan:2012ip, Wall:2012uf, Headrick:2014cta}. 
In this paper we provide 
further support by matching results on the two sides of the duality in a class of excited 
states, and in the process develop new tools to analyze universal behavior in large-$N$ 2d CFTs.

The class of CFT states we consider are excited above the vacuum by the insertion of a 
primary operator at a point. The general problem of Renyi and von Neumann entanglement entropies of an interval in such states was considered in 
\cite{2011PhRvL.106t1601A, 2012JSMTE..01..016I,
Mosaffa:2012mz}, 
which found the leading, universal corrections to the vacuum values in an expansion in the interval length. 
These corrections conform to a so-called first law of entanglement thermodynamics \cite{Bhattacharya:2012mi, Allahbakhshi:2013rda}. A holographic explanation of these terms can be found in \cite{Astaneh:2013gp}. Excited states in symmetric orbifolds, and the D1D5 CFT in particular, were studied in \cite{Asplund:2011cq, Giusto:2014aba}, and further progress was made for rational CFTs in \cite{Nozaki:2014hna,He:2014mwa,Nozaki:2014uaa, Palmai:2014jqa}.

In a theory with large $N$, and therefore
large central charge $c$, we can distinguish between light operators with dimension $\Delta \ll c$ and heavy operators with $\Delta =  O(c)$.  States produced by light primaries in large-$c$ CFTs were considered in \cite{Caputa:2014vaa}.  Here we will consider heavy states, which are dual to geometries with non-trivial backreaction of the metric. Since the holographic dual is three-dimensional and there is no propagating graviton, the backreaction is limited to local defects and global properties (deficit angles and black holes). Given certain assumptions about the operator-product expansion in large-$c$ CFTs, we find precise agreement with holographic computations of the entanglement entropy for a single interval in an excited state.  Our analysis is in the same spirit as 
the study of entanglement entropy in \cite{tom} and relies on large-$c$ Virasoro 
conformal blocks obtained in \cite{Fitzpatrick:2014vua}.

We study Renyi entropy in these states using the replica trick,
described in sec.~\ref{ss:replica}. This allows us to compute the Renyi 
entropy of a single interval in terms of a four-point function of two heavy operators, which excite the state, 
and two twist operators. Using the OPE, the four-point function can be expanded in conformal blocks. 
We restrict to states with no macrosopic $O(c)$ expectation values for light operators, and assume that in CFTs with a large central charge and sparse spectrum the expansion is
well approximated by the identity block. This was proved for a different type of twist correlator (corresponding to the torus partition function) in sparse CFTs in \cite{hks}, but here it is just a conjecture. In general, the identity contribution is not known in its full 
analytic form, though it can be computed in a small-interval expansion. However, in the limit relevant for the entanglement 
entropy, the twist operators become light and 
the corresponding `light-light-heavy-heavy' vacuum block has been computed in the large-$c$ limit in \cite{Fitzpatrick:2014vua}.
The entanglement entropy then takes a simple closed form, given in sec.~\ref{ss:blocke}. 

The approximation of the full four-point function by the identity contribution is not single valued on the complex plane; 
different ways of analytically continuing it define multiple OPE channels.  
We show in sec.~\ref{ss:blockgeo} that the semiclassical conformal block is directly related to the length of geodesics in a Euclidean asymptotically AdS$_3$ geometry with a defect. Different channels compute the length of geodesics with different winding around the defect. 
The correlator, and thus the entanglement entropy, is given by the identity block in the dominant channel, which corresponds 
to the geodesic of minimal length.

This CFT result agrees precisely with holographic gravity calculations.
Examples of the heavy states that we consider include high-energy eigenstates of the Hamiltonian dual to conical defects and microstates of BTZ black holes, discussed in sec.~\ref{s:staticex}. We find that the single interval entanglement entropy in these high-energy pure states on a circle behaves like that of a finite temperature CFT on a line.
 
An interesting dynamical setup related by conformal mapping and analytic continuation is 
that of a local quench, dual to a highly boosted black hole or conical defect. 
In sec.~\ref{s:loq}, we show that the evolution of the single interval entanglement entropy precisely matches the holographic computation of \cite{Nozaki:2013wia}.
This is similar to what was found in \cite{Hartman:2013qma}, where the authors proposed an 
exact holographic dual to a global quench of a 2d CFT, and verified that the 
evolution of the single interval entanglement entropy was correctly reproduced. 
Our findings provide new support to the holographic covariant proposal for computing the entanglement entropy.
We conclude in sec.~\ref{ss:cc} by commenting on the relation of these local quenches
to the well-known ones studied in 
\cite{Calabrese:2007,Stephan:2011}, and holographically in 
\cite{Nozaki:2013wia, Ugajin:2013xxa, Asplund:2013zba}.

\section{Twist correlators in sparse CFTs}\label{s:twist}
\subsection{Review of the replica trick}\label{ss:replica}
The entanglement entropy of a geometric region $A$ is the von Neumann entropy of the reduced density matrix $\rho_A = \tr_{A^c}\rho$,
\be
S_A = - \tr \rho_A \log \rho_A \ .
\ee
Often it is easier to compute the associated Renyi entropies,
\be
S_A^{(n)} = \frac{1}{1-n}\log \tr \rho_A^n
\ee
for integers $n\geq 2$, and analytically continue $n \to 1$ to recover the entanglement entropy.   In particular, if the state associated to the density matrix  $\rho$ can be prepared by a Euclidean path integral, then the Renyi entropy can be computed by a path integral for multiple copies of the system glued together along region $A$ (see  \cite{Calabrese:2009qy} for a  review). States that meet this criteria include the vacuum, prepared by a path integral on a half-plane or disk; thermal states, prepared by a path integral on a cylinder or torus; and states obtained from these by acting with operator insertions.
\begin{figure}
\begin{overpic}[scale=1]{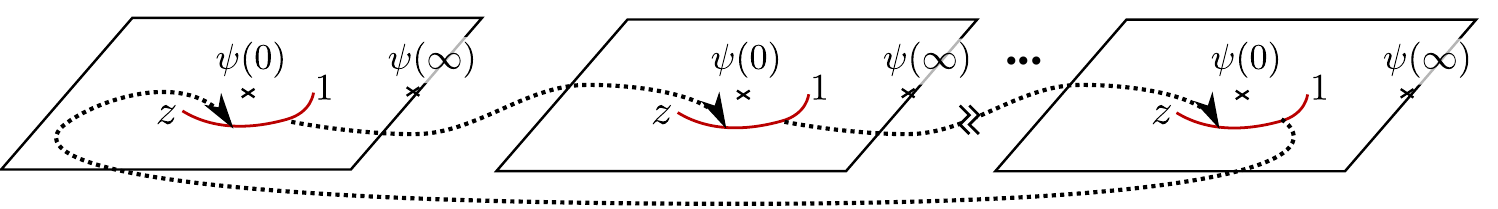}
\end{overpic}
\caption{Configuration of $2n$ $\psi$'s on a multisheeted surface, branched across the red cut which extends along an arc on the unit circle from 1 to $z$.\label{fig:psic}}
\end{figure}

A general pure state in CFT can be created by inserting a local operator at the origin, by the state-operator correspondence \cite{Polchinski:1998rq}.  The focus of this paper is on primary states and their conformal descendants. Primary states are pure states produced by inserting a local primary operator,
\be
|\psi \rangle = \psi(0) |0\rangle \ , \qquad \langle \psi | = \lim_{z,\bz \to \infty} \bz^{2h_\psi}z^{2\bh_\psi}\langle 0 | \psi(z,\bz) \ ,
\ee 
where $\psi$ has dimensions $(L_0, \overline{L}_0) = (h_\psi ,\bh_\psi)$. For now, region $A$ is taken to be a segment of the unit circle in the complex plane, extending from 1 to $z$. The replica partition function in the state $\rho(\psi) \equiv |\psi\rangle\langle \psi |$ is computed by a path integral on $n$ copies of the system glued together along $A$,
\be\label{gf}
G_n(z,\bz) \equiv \tr \rho(\psi)_A^n = \langle \psi(0_1)\psi(\infty_1) \psi(0_2)\psi(\infty_2) \cdots \psi(0_n)\psi(\infty_n)\rangle_{\Sigma_n} \ ,
\ee
where $\Sigma_n$ is an $n$-sheeted manifold branched along $A$ as depicted in fig.~\ref{fig:psic}, and the subscripts label the sheets on which operators are inserted. Instead of viewing this as a correlation function on a multisheeted surface we can instead view it as a correlator including twist operators,
\bea\label{gcor}
G_n(z,\bz) &=& \langle \Psi | \sigma_n(z,\bz)  \tsigma_n(1) |\Psi \rangle  \notag\\
&=& \langle \Psi(\infty) \sigma_n(z,\bz) \tsigma_n(1) \Psi(0)\rangle \ .
\eea
We have ignored a UV-sensitive constant, coming from regulating the twist operators, which will be restored below.  
This is a 4-point function of local operators, not in the original CFT but in the cyclic orbifold theory $CFT^n/Z_n$. 
The twist operators $\sigma$ and $\tsigma$ (which have opposite orientation) have dimension
\be
H_n = \bH_n = \frac{c}{24}\(n-\frac{1}{n}\) ,
\ee
and $|\Psi\rangle$ is the state in the orbifold theory obtained by inserting $\psi$ in all $n$ copies, 
\be
\Psi = \psi_1 \psi_2\dots \psi_n \ , 
\ee
where the subscripts indicate different copies of the CFT.
\subsection{Conformal block expansion}\label{ss:blocke}

Using the OPE, any 4-point function including (\ref{gcor}) can be expanded in conformal blocks. Schematically,
\be\langle \psi^n \sigma_n \tsigma_n \psi^n \rangle = \sum_{\mbox{primaries $p$}}a_p \quad  \raisebox{-0.5\height}{
\begin{overpic}{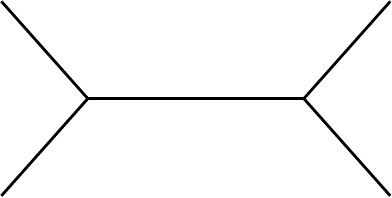}
\put(-70,-10) {$\sigma$}
\put(-70,500) {$\tsigma$}
\put (450,290) {$\mathcal{O}_p$}
\put (1050,-10) {$\psi^n$}
\put (1050,500) {$\psi^n$}
\end{overpic}
} \qquad .
\ee
In more detail,
\be\label{gexp}
G_n(z,\bz) = \sum_p a_p \bF(c_n, h_p, h_i, 1-z)\bF(c_n, \bh_p, \bh_i, 1-\bz) \ ,
\ee
where $\mathcal{F}$ is the Virasoro conformal block, a function fixed entirely by the conformal algebra.
The sum converges for $|z-1|<1$. In this expression
\be
c_n \equiv c n
\ee
is the central charge of the orbifold $CFT^n/Z_n$, the sum runs over Virasoro primary operators $\bO_p$ of dimension $(h_p, \bh_p)$ in the orbifold theory, $a_p$ is a constant related to OPE coefficients, and $h_i = h_{1,2,3,4}$ denotes the external weights. In our case we have two external operators that create the state and two twist operators,
\be\label{ourweights}
h_1 = h_4 = h_\Psi = n h_\psi , \qquad h_2 = h_3 = H_n \ .
\ee
We have expanded in the $t$ channel $z \to 1$, which corresponds to bringing together $\sigma_n$ and $\tsigma_n$.\footnote{This choice of channel is reflected in the fact that the argument of the conformal block $\bF$ is $1-z$. Our convention is that $\bF(\dots, z)$ is the ordinary $s$-channel block.}  The conformal block $\bF$ is not known in closed form, but it is fixed by the Virasoro algebra and can be expanded in a power series,
\be\label{bigfe}
\bF(c, h_p, h_i, z) = z^{h_p - h_1 - h_2}\left(1 +  \frac{(h_p+h_2-h_1)(h_p+h_3-h_4)}{2h_p}z + \cdots\right) \ .
\ee

We now specialize to the class of CFTs relevant for holography.  These are CFTs with large central charge $c$ and a sparse spectrum of low-dimension operators. 
It is known that in the limit $c \to \infty$ with $h_p/c$ and $h_i/c$ held fixed, the conformal block exponentiates \cite{Zamolodchikov:1987}:
\be
\bF(c, h_p, h_i, z) \approx \exp\left[ -\frac{c}{6}f\left(\frac{h_p}{c}, \frac{h_i}{c}, z\right)\right] \ .
\ee
Since (\ref{gexp}) becomes a sum of exponentials, we expect that under certain circumstances it is dominated by the largest term.  For $z\to 1$, the largest term must be the contribution of the identity representation $\bO_p =\mathds{1}$, for which $a_p=1$,  so:
\be\label{gapp}
G_n(z,\bz) \approx \exp\left[ -\frac{c_n}{6}f_0(h_i/c_n, 1-z) - \frac{c_n}{6}f_0(\bh_i/c_n, 1-\bz)\right]
\ee
where\footnote{It was shown in \cite{tom} that the large-$c$ limit commutes with the limit $h_p/c \to 0$, so this expression is well defined.} 
\be\label{fzz}
f_0(h_i/c, x) \equiv f(0, h_i/c, x) \ .
\ee
This is the contribution from the identity and all of its descendants, which consist of all operators constructed from the stress tensor and its derivatives. We will not attempt to define precisely in what class of theories or in what range of $z$ the approximation (\ref{gapp}) is valid.  We expect it to hold for a finite range of $z$ in theories with a large central charge and a sparse spectrum of light ($\Delta \ll c$) operators.\footnote{The definition of `sparse' suitable to reproduce the thermodynamics of 3d gravity is that the density of states is bounded by $\rho(\Delta) \lesssim e^{2\pi \Delta}$ \cite{hks}; this is likely to be at least a necessary criterion here as well.}

 If there are light operators besides the stress tensor, then we will also restrict to states $|\Psi\rangle$ which do not have macroscopic, $O(c)$, expectation values for these light operators.  Otherwise, we expect large OPE coefficients $a_p$ in (\ref{gexp}). For $\bO_p$ constructed from a product of $n$ light operators with large expectation values, $a_p$ is expected to have a contribution proportional to $\langle \Psi | \bO_p | \Psi\rangle \sim c^n$, and may therefore contribute to leading order in the $1/c$ expansion. In fact, similar contributions have been found explicitly in supersymmetric CFTs \cite{Giusto:2014aba}.\footnote{We thank the authors of \cite{Giusto:2014aba} for bringing this to our attention.}

These are all necessary conditions, but we have not proved they are sufficient; further restrictions on the growth of OPE coefficients may also be required. For the rest of the paper we will simply assume that (\ref{gapp}) is a good approximation, up to corrections non-perturbative in the $1/c$ expansion.

The universal contribution (\ref{gapp}) is then a prediction for the replica partition function with primary operator insertions in sparse CFTs. It can be easily expanded around $z\sim 1$ to any desired order using standard techniques to compute conformal blocks (see appendix A of \cite{tom} for a review):
\bea\label{gseries}
\log G_n(z,\bz) &=& - 2 c_n \delta_n \log (1-z)\  \\
&&  -\frac{c_n }{6} \left[  a_2 (1-z)^2 +  a_3 (1-z)^3 + a_4 (1-z)^4+ O(1-z)^5  \right]   \notag \\
&& +\  (z \to \bz \ , \ \delta_\psi \to \bdelta_\psi)   \ ,  \notag
\eea
where
\be
a_2 = a_3 = - 12 \delta_\psi \delta_n \ , \qquad
a_4=\frac{6}{5} \delta_\psi \delta_n \(-9 - 2 (\delta_\psi + \delta_n) + 44 \delta_\psi \delta_n   \) \,  , 
\ee
and
\be
\delta_\psi = \frac{h_\Psi}{c_n} = \frac{h_\psi}{c} \ , \quad \bdelta_\psi = \frac{\bh_\psi}{c} , \quad \delta_n = \frac{1}{24}\(1-\frac{1}{n^2}\)
\ .
\ee

For the entanglement entropy, $n \to 1$, the answer can be obtained in closed form without resorting to a series expansion.  In this limit, the dimension of the twist operators goes to zero, so we need the identity block where two external operators are heavy and two are light.  This was computed in the large-$c$ limit in \cite{Fitzpatrick:2014vua} using the monodromy method for the Virasoro blocks \cite{Zamolodchikov:1987}. This method, reviewed in \cite{tom,Fitzpatrick:2014vua}, translates the computation of $f\( h_p/c, h_i/c, z\)$ into the problem of fixing the monodromy for a second order differential equation related to the conformal Ward identity. In \cite{Fitzpatrick:2014vua}, this was achieved to linear order in a perturbative expansion in $h_2/c = h_3/c$ for a light state running in the intermediate channel.

Denoting this `light-light-heavy-heavy' identity block by
\be
g(h_\psi, \epsilon; z) \equiv f_0(h_i/c, z) , \quad \mbox{with} \quad h_1=h_4 = h_\psi , \quad h_2=h_3 = \epsilon \frac{c}{24}
\ee
the result to leading order in $\epsilon$ is \cite{Fitzpatrick:2014vua}
\be\label{fpsi}
g(h_\psi, \epsilon; 1-z) = \frac{\epsilon}{2} \log\left( \frac{1-z^{\alpha_\psi}}{\alpha_\psi}\right) + \frac{\epsilon}{4}(1-\alpha_\psi)\log z
\ee
where
\be
\label{eq:alpha_psi}
\alpha_\psi \equiv \sqrt{1 - 24 h_\psi/c} \ .
\ee
Twist operators have $\epsilon \sim 2(n-1)$ as $n \to 1$, so
\be\label{ftc}
\log G_n(z,\bz) = \frac{c}{6}(1-n)\log \left[\frac{z^{ \frac{1}{2}(1-\alpha_\psi)}\bz^{ \frac{1}{2}(1-\balpha_\psi)}(1-z^{\alpha_\psi})(1-\bz^{\balpha_\psi})}{\alpha_\psi\balpha_\psi}\right]+ O((n-1)^2) \ .
\ee
This is the main technical formula that will be used to compute various entanglement entropies throughout the paper.

\subsection{Conformal block as geodesic length}\label{ss:blockgeo}

The light-light-heavy-heavy conformal block is directly related to the length of a  geodesic on a 3d geometry with a defect.  The other comparisons between correlators and geodesic lengths in this paper all follow from this one by acting with conformal transformations and the appropriate analytic continuations.  Consider the asymptotically Euclidean  AdS$_3$ geometry
\be\label{defge}
ds^2 = \frac{L}{2}dz^2 + \frac{\bL}{2}d\bar z^2 + \left(\frac{1}{y^2} + \frac{y^2}{4}L\bL\right) dz d\bz +  \frac{dy^2}{y^2} \ , 
\ee
with
\be
L = L(z) \ , \qquad \bL = \bL(\bz) \ ,
\ee
where $z$ and $\bz$ are complex coordinates on the boundary plane, $y$ is a bulk coordinate and we have set the radius of AdS to unity. 
This is a solution of the vacuum Einstein equations for any $L$, $\bL$ (away from possible singularities), and in such a state the holographic dictionary relates $L$ with the CFT stress tensor $T\equiv \langle T_{zz}(z)\rangle $ as 
\be
T =  -\frac{c}{12 } L\ .
\ee
 Let us choose
\be\label{tprim}
T = \frac{ h_\psi }{z^2} \ , \qquad \bT = \frac{ \bh_\psi}{\bz^2} \ .
\ee
This is the semiclassical stress tensor in the presence of heavy operator insertions at $z = 0, \infty$, both with weights $(h_\psi, \bh_\psi)$. The same stress tensor was used in the CFT derivation of the light-light-heavy-heavy Virasoro block \cite{Fitzpatrick:2014vua}. Clearly the geometry is singular at $z = 0$.

The length of a geodesic $\gamma$ in this geometry, anchored near the boundary at 
\be \label{endpoints}
(y, z) = \( \epsilon_{\text UV}, 1 \) \ , \qquad (y,z) = \( \epsilon_{\text UV}, z_0\)
\ee
is straightforward to calculate. One can, for example, take the expression for geodesics in pure AdS$_3$ and then pull it back to the geometry \eqref{defge} (see appendix \ref{aa:geod}). The answer is related to the semiclassical conformal block  by 
\be\label{geomatch}
{\cal L}_\gamma = \frac{2}{\epsilon}\left( g(h_\psi, \epsilon; 1-z_0) + g(\bh_\psi, \epsilon; 1-\bz_0) \right) - 2\log \epsilon_{\text UV}\, ,
\ee
where $g$ was given in (\ref{fpsi}). 

There are different geodesics with the same endpoints but that differ in their winding around the singularity at $z=0$. All of their lengths are captured by \eqref{geomatch}. The choice of branch cut 
for $z_0^{\alpha_\psi}$ in this formula selects a winding for the geodesic around the singularity.

\begin{figure}
\centering
\includegraphics[width=\textwidth]{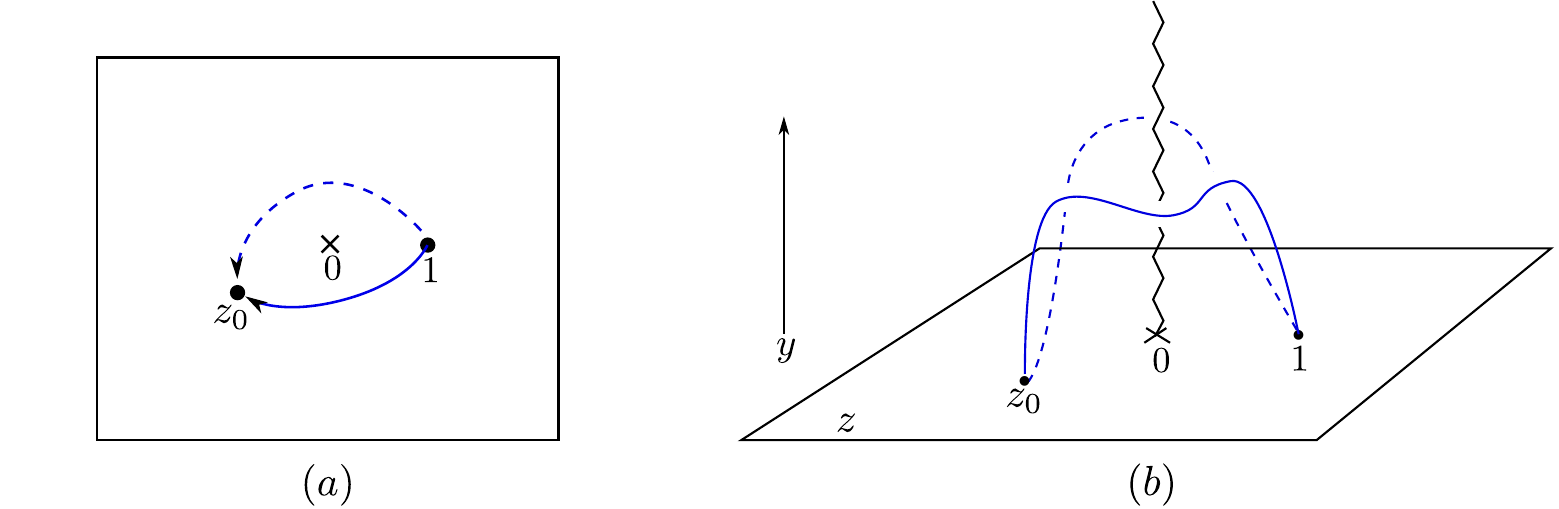}
\caption{\textit{Left:} Two ways to analytically continue the approximate expression for the twist correlator $G_n(z,\bz)$ around the singularity in the complex $z$-plane. \textit{Right:} Two geodesics in the singular geometry \eqref{defge} with the same endpoints. The choice of analytic continuation in evaluating the block translates into a choice of winding around the singularity at $z= 0$.\label{fig:defectwinding} }
\end{figure}

This has a direct counterpart in our CFT computation. When we approximate the full Euclidean correlator $G_n(z,\bz)$ by the identity block in (\ref{gapp}) we must choose a way of analytically continuing around the singularity at $z = 0$. This would not be necessary if we knew the exact correlator, which is single valued on the complex plane, but it is required once we replace the full correlator by its identity-block approximation since $\bF(z)$ is not single valued. Taking the analytic continuation around $z = 0$ along the solid curve  or  along the dashed curve in fig.~\ref{fig:defectwinding}a defines  alternative conformal block expansions. 

In approximating the twist correlator $G_n(z,\bz)$ by the identity block we therefore need to consider all these possible channels.
Since we are assuming that the correlator is dominated by the identity block in sparse CFTs, the full answer is the identity block in the dominant channel, which is equal to the minimal geodesic length. Non-minimal geodesics (which may play an important role in reconstructing bulk geometry from entanglement \cite{Myers:2014jia,Hubeny:2014qwa,Czech:2014wka,Balasubramanian:2014sra,Headrick:2014eia,Czech:2014ppa}) correspond to the identity block in subdominant OPE channels.

In  \cite{Fitzpatrick:2014vua} the semiclassical block $g$ was also interpreted in terms of the motion of particles on a defect geometry.  It was shown that $g$ encodes information about the energy shift of bound states on the defect background, and about the spectrum of quasinormal modes on BTZ.  Our interpretation of the semiclassical block as a geodesic length in the geometry (\ref{defge}) is of course closely related. Indeed, large mass quasinormal modes can be computed by geodesics in the WKB approximation (see for example \cite{Fidkowski:2003nf}). In a general large-$c$ CFT, the match to bulk results discussed in \cite{Fitzpatrick:2014vua} holds only at very high energies, but in sparse CFTs where our approximations apply for all $z$, many of the results of \cite{Fitzpatrick:2014vua} should hold at all frequencies. 

\section{Black holes and conical defects}\label{s:staticex}

\subsection{Entanglement entropy on a circle}\label{ss:soncircle}
As the first application of (\ref{ftc}) we consider a CFT on a circle $w \sim w + 2\pi$ and let region $A$ be a segment of length $\ell$.  This is related to the coordinates of section \ref{s:twist} by mapping the cylinder to the plane, $z = e^{i w}$.  The twist operators are located at $w = 0$ and $w = \ell$.  Therefore we take
\be\label{cfgz}
z = e^{i\ell} , \qquad \bz = e^{-i\ell}
\ee
in (\ref{ftc}) to find the entanglement entropy of region $A$. For a state with zero spin, $h_\psi = \bh_\psi$, the result is
\footnote{
We understand some of these results have been obtained independently by \cite{Takayanagi}.
}
\be\label{scircle}
S_A = \frac{c}{3}\log\left[ \frac{\beta_\psi}{\pi \eps_{UV}}\sinh\left(\frac{\ell \pi}{\beta_\psi}\right)\right]
\ee
where
\be\label{defb}
\beta_\psi \equiv \frac{2\pi}{\sqrt{24 h_\psi/c - 1}} \ .
\ee
We have inserted a UV cutoff, $\eps_{UV}$, to regulate the twist operators \cite{Calabrese:2009qy}.

The interpretation of this formula depends on whether $\beta_\psi$ is real or imaginary.  We first discuss the case of real $\beta_\psi$, \ie $h_\psi > \frac{c}{24}$. The entanglement entropy (\ref{scircle}) is the well known result for a CFT on a line at inverse temperature $\beta_\psi$ \cite{Holzhey:1994we,Calabrese:2004eu}, and furthermore, the temperature relation (\ref{defb}) is precisely the transformation from the microcanonical to the canonical ensemble in sparse CFTs \cite{hks}.  The fact that we have reproduced this formula here is nontrivial, for two reasons: first, our CFT is on a circle rather than a line, and second, our CFT is in a pure state $\rho = |\psi \rangle \langle \psi |$ rather than a thermal state.  

The entanglement entropy (\ref{scircle}) agrees precisely with the holographic entanglement entropy formula applied to a BTZ black hole at temperature $\beta_\psi$ \cite{Ryu:2006bv, Ryu:2006ef}.  The map (\ref{defb}) is the usual relation between temperature and energy for such a black hole.  

\begin{figure}
\centering
\includegraphics{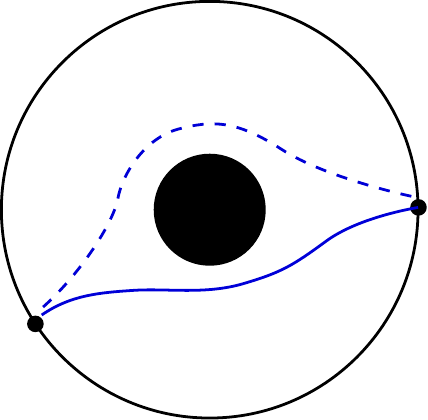}
\caption{ Two geodesics in the BTZ geometry (at fixed time) with the same endpoints.  The choice of channel in the CFT corresponds to a choice of winding around the black hole horizon. \label{fig:curves}}
\end{figure}

Our CFT calculation corresponds to a particular microstate of BTZ. This raises a potential puzzle, because in a pure state the entanglement entropy of region $A$ is equal to the entanglement entropy of its complement $A^c$, whereas our formula as written is not invariant under $\ell \to 2\pi - \ell$.  
The resolution of this puzzle comes from considering different channels for the OPE approximation.  As explained in section \ref{ss:blockgeo}, we should consider the correlator as given  by the identity block in the dominant channel. 
In the formula (\ref{scircle}) we implicitly used the analytic continuation along the dashed curve in fig.~\ref{fig:defectwinding}a, but we could have taken $z$ along the solid curve instead, which amounts to replacing  $\ell \to 2\pi - \ell$ in (\ref{scircle}). Multiple windings are also possible, but always subdominant. Therefore we should replace $\ell \to \min(\ell, 2\pi - \ell)$ in (\ref{scircle}) for the full answer.
 At $\ell = \pi$, the two expansions exchange dominance and $S_A$ has a discontinuous first derivative.

In the BTZ black hole, these two OPE expansions correspond to two different geodesics, shown in fig.~\ref{fig:curves}.  There is also a third geodesic on BTZ that is important in the holographic entanglement entropy calculation at finite temperature but did not make any appearance in our CFT calculation: the disconnected geodesic consisting of the solid curve in fig.~\ref{fig:curves} plus another curve wrapping the horizon.  This disconnected geodesic gives the correct holographic entanglement entropy for a region of length $\pi < \ell < 2\pi$ in a thermal state. The horizon segment is required to satisfy the homology condition which is part of the holographic prescription, and accounts for the fact that $S_A \neq S_{A^c}$ in a mixed state \cite{Headrick:2007km}. Since we are working in a pure microstate, we do not expect such a contribution, and indeed we find that our CFT calculation agrees with the holographic calculation \textit{without} imposing the homology condition.\footnote{Another difference between our pure-state calculation and the finite-temperature calculation is that we do not (and should not) see the Hawking-Page transition. At finite temperature, holographic entanglement entropy should be computed in the geometry that dominates the canonical ensemble. At high temperature $\beta_\psi < 2\pi$ this is BTZ, but below the Hawking-Page transition at $\beta_\psi = 2\pi$, \ie $h_\psi = \frac{c}{12}$ the dominant geometry is thermal AdS \cite{Maldacena:1998bw}.  In our pure-state calculation we have fixed the energy, not the temperature, so the result agrees with the black hole geometry even at energies below the transition.}

Finally we turn briefly to the case $h_\psi < \frac{c}{24}$, where $\beta_\psi$ is imaginary.  The dual geometry is a conical defect in the center of global AdS.  The formula (\ref{scircle}) agrees once again with the holographic entanglement entropy.

These matches between entanglement entropy and geodesic lengths follow from the general relationship between conformal blocks and geodesics discussed in section \ref{ss:blockgeo}. Although the defect geometry computes the correct geodesic lengths, we do not mean to imply that the defect geometry is precisely dual to a CFT microstate. The geometry is singular, and therefore sensitive to UV effects in quantum gravity, and furthermore different CFT microstates with the same values of $(h_\psi, \bh_\psi)$ map to the same bulk defect geometry.  Therefore the more accurate statement is that the defect geometry captures certain features of the microstate, including the entanglement entropy. The true bulk microstate need not be geometric at all, and to describe it in detail would presumably require string theory.

Certain microstates in supersymmetric field theories dual to string theory were considered in \cite{Giusto:2014aba}, where the authors found non-universal (state-dependent) contributions to the entanglement entropy.  These are microstates constructed from a superposition of Ramond groundstates designed to produce an atypically large expectation value for light supergravity fields \cite{Kanitscheider:2007wq}.  Generally we do not expect our results to apply to such states for the reasons discussed below (\ref{fzz}) --- our restriction on the growth of OPE coefficients ensures that only the gravity sector of the bulk theory is important.

\subsection{Renyi entropies}

The Renyi entropy computed from the series expansion (\ref{gseries}) with $z = e^{i\ell}$, $\bz = e^{-i\ell}$, is
\bea\label{snf}
S_A^{(n)} &=& \frac{1}{1-n}\log G_n(z,\bz) \notag \\
&=&  \frac{c}{6} \frac{(1+n)}{n}  \left[  \log \ell -\frac{\ell^2}{24} -\frac{\ell^4}{2880} + \delta_\psi   \ell^2 + \frac{ \delta_\psi}{60}  (1 - 12 \delta_\psi )\ell^4 \right]  \\
& &-\frac{c}{6} \frac{(1+n)(n^2-1)}{n^3}   \frac{ \delta_\psi  - 22 \delta_\psi^2 }{120}  \ell^4   + O(\ell^5) \, , \notag 
\eea
where $\delta_\psi = h_\psi/c$. 

This is a CFT prediction for the Einstein action evaluated on the bulk replica manifold obtained by solving the Einstein equations with boundary conditions as in fig.~\ref{fig:psic}. Although we have not done so, it is likely that the methods of \cite{Faulkner:2013yia} could be used to construct this geometry explicitly. If so, then the Einstein action would be determined by the same monodromy prescription used to compute the Virasoro block in \cite{Fitzpatrick:2014vua}, and therefore would manifestly agree with (\ref{snf}). This would be a very direct relationship between the CFT calculation and the bulk geometry, analogous to the relationship for two intervals in vacuum demonstrated in \cite{tom,Faulkner:2013yia}.

Our result (\ref{snf}) is the Renyi entropy in a pure state on a circle in a small $\ell$ expansion at fixed $\delta_\psi$.  Renyi entropies on a circle at finite temperature have been computed holographically in \cite{Barrella:2013wja} and in CFT in \cite{Chen:2014unl}, using a high-temperature expansion.
It would be interesting to repeat our microstate calculation in the corresponding high-energy limit, which is defined by $\delta_\psi \to\infty$ with $\ell \delta_\psi $ held fixed, in order to compare to the thermal result.  The full Renyi entropy in a thermal state cannot equal the Renyi entropy in a pure state, since the former is not invariant under $\ell \to 2\pi - \ell$ \cite{Barrella:2013wja}. However, this does not rule out the possibility that the $O(c)$ contribution can agree.

\subsection{Angular potential}

The discussion of the previous sections immediately generalizes to excited states with $h_\psi \neq \bh_\psi$. 
Taking $z = e^{i \ell}, \bz= e^{-i \ell}$ in (\ref{ftc}) and generic  $(h_\psi, \bh_\psi)$ we find the entanglement entropy 
\be\label{sang}
S_A = \frac{c}{6}\log\left[ \frac{\beta_\psi\bbeta_\psi}{\pi^2 \eps^2_{UV}}\sinh\left(\frac{\min(\ell, 2 \pi -\ell) \pi}{\beta_\psi}\right) \sinh\left(\frac{\min(\ell, 2 \pi -\ell) \pi}{\bbeta_\psi}\right)\right] \, ,
\ee
where now
\be\label{defbb}
\beta_\psi \equiv \frac{2\pi}{\sqrt{24 h_\psi/c - 1}} \ , \qquad \bbeta_\psi \equiv \frac{2\pi}{\sqrt{24 \overline{h}_\psi/c - 1}} \ .
\ee

For real inverse temperatures, the entanglement entropy (\ref{sang}) matches  the entanglement entropy for a CFT on a line at inverse temperature $\beta$ and angular potential $\Omega$   \cite{Hubeny:2007xt}. The relation to the effective temperatures for  left and right moving modes is 
\be
\beta_\psi = \beta (1 + \Omega) \ , \qquad  \bbeta_\psi= \beta (1 - \Omega) \, .
\ee
The result \eqref{sang} agrees with the holographic computation of the entanglement entropy  of a rotating BTZ black hole \cite{Hubeny:2007xt} (without the homology condition).

\section{Local operator quenches}
\label{s:loq}

So far we have restricted to static configurations and computed geodesic lengths in Euclidean AdS$_3$.  We now turn to the time-dependent entanglement after a local quantum quench, for which the holographic calculation involves geodesics in Lorentzian signature.  

A quench is a sudden change in the system that produces a time-dependent excited state.  To model this process, we consider a state on the real line created by the insertion of a local operator at $x = 0$, $ t = i \delta$.  The offset into the imaginary time direction is necessary to produce a normalizable state.  Physically this is quite different from the primary states on a circle considered above, though the calculation will turn out to be related by conformal symmetry.  For the local quench, the operator is inserted close to the line on which the state is defined, so it creates a localized excitation.  This excitation then spreads out over the system, so we expect corresponding behavior for the entanglement entropy.

This model for a quench was recently analyzed in \cite{Nozaki:2014hna,He:2014mwa,Nozaki:2014uaa} in the context of rational CFTs, and for large-$c$ CFTs in \cite{Caputa:2014vaa}. The results in \cite{Caputa:2014vaa} apply to quenches produced by light operators, which do not backreact in the bulk.  We consider heavy operators, with dimension $O(c)$.

Note that a different type of quench was considered by Calabrese and Cardy in \cite{Calabrese:2007,Calabrese:2009qy}. The relation between the two is discussed in more detail below.  

\subsection{Identity block approximation}

\begin{figure}
\centering
\begin{overpic}[scale=1]{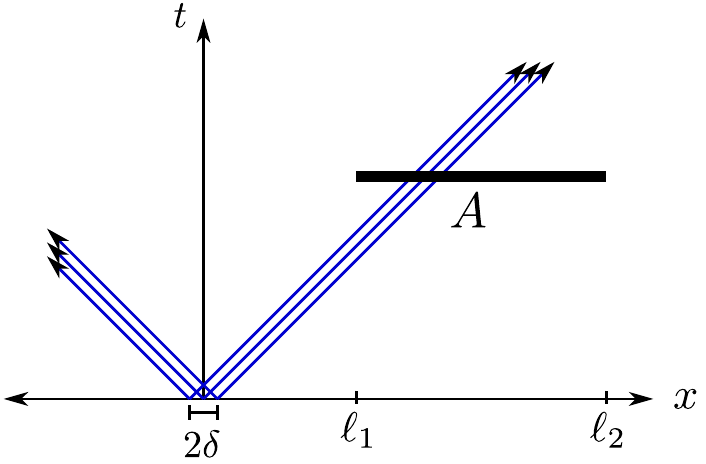}
\end{overpic}
\caption{Setup for the local operator quench. The initial state has a localized excitation near $x = 0$, which propagates outward and eventually increases the entanglement entropy of region $A$.
\label{fig:quenchsetup}}
\end{figure}

Following \cite{Nozaki:2014hna,He:2014mwa,Nozaki:2014uaa,Caputa:2014vaa}, the real-time density matrix after a local operator quench is
\be
\rho(t) = N e^{-iHt}\psi(w_4, \bw_4) |0 \rangle \langle 0 | \psi(w_1, \bw_1)e^{iHt}
\ee              
where the normalization factor $N$ is fixed by $\tr \rho(t) = 1$, and
\be
w_4 = i\delta  \ , \quad \bw_4 = -i \delta  , \quad w_1 = -i \delta  , \quad \bw_1 = i \delta \ .
\ee
We assume $h_\psi = \bh_\psi$. The quench is local in the limit $\delta \to 0$ but we will take $\delta$ finite for now.
We choose region $A$ to be the interval $[\ell_1, \ell_2]$ at time $t$, with $|\ell_1| < |\ell_2|$. See fig.~\ref{fig:quenchsetup} for the case $\ell_1>0$.
The Renyi entropy can be computed from the two-point correlator of twist operators in this excited state, 
\be \label{eq:4pt}
A_n(w_i, \bw_i) \equiv \langle \Psi(w_1, \bw_1) \sigma_n(w_2, \bw_2) \tsigma_{n} (w_3, \bw_3) \Psi(w_4, \bw_4) \rangle\ .
\ee
This is a correlation function in the cyclic orbifold $CFT^n/Z_n$, with the twist operators inserted at the interval endpoints,
\be
 w_2 = \ell_1 +t \ , \quad \bw_2 = \ell_1 - t \ , \quad w_3 = \ell_2 + t \ , \quad \bw_3 = \ell_2 - t \ .
\ee
From \eqref{eq:4pt}, we can compute\footnote{$w_{ij} \equiv w_i - w_j$ and $|w|^2 \equiv w \bw$.}
\be
 \Tr \rho(t)^n  = \frac{A_n(w_i,\bw_i) }{\(\langle \psi(w_1, \bw_1)  \psi(w_4, \bw_4)  \rangle\)^n} =|w_{14}|^{4 n h_\psi} A_n(w_i,\bar w_i)  \ .
\ee
Applying a global conformal transformation 
\be \label{conftr}
z(w) = \frac{(w_1-w)w_{34}}{(w-w_4)w_{13}} \ ,
\ee
this becomes
\be
\Tr \rho(t)^n  = |w_{23}|^{-4nH_n}|1-z|^{4nH_n}G_n(z, \bz) 
 \ee
where $G_n$ is the 4-point function analyzed in section \ref{ss:blocke}, as a function of the cross-ratio 
\be
z \equiv \frac{w_{12} w_{34}}{w_{13} w_{24}} = \frac{ (\ell_1+t+i\delta)(\ell_2+t-i\delta)}{(\ell_1+t-i\delta)(\ell_2+t+i\delta)} \ .
\ee
For small $\delta$,
\be\label{zcs}
z = 1 + \frac{2i(\ell_2-\ell_1)}{(\ell_1+t)(\ell_2+t)}\delta +O(\delta^2) \ , \qquad
\bz = 1 - \frac{2i(\ell_2-\ell_1)}{(\ell_1-t)(\ell_2-t)}\delta +O(\delta^2) \ .
\ee
As discussed in section \ref{ss:blocke}, in CFTs with a large central charge and a sparse spectrum of low-dimension operators, we expect that $G_n(z,\bz)$ can be approximated by the contribution from the identity and its Virasoro descendants, and that corrections are exponentially suppressed in the $1/c$ expansion.  Thus the Renyi entropy is
\be\label{snfq}
S^{(n)}_{A} = \frac{1}{1-n}\log \left[  |w_{23}|^{-4nH_n} |1-z|^{ 4n H_n}\left|\exp\left(-\frac{c_n}{6}f_0(h_i/c_n, 1-z)\right)\right|^2 \right] \ .
\ee
In principle this is the final answer for the Renyi entropy.  At early times, it can be evaluated as a series expansion around $z \sim 1$ using the explicit Virasoro block (\ref{bigfe}).  We will assume (\ref{snfq}) holds for all times in a sparse CFT, but for $|t|>|\ell_1|$ the usual series expansion is invalid  (even though $z,\bz \sim 1$) due to branch cuts in the Lorentzian correlator as discussed below.

As $n \to 1$, the twist operators effectively become light, and plugging in (\ref{ftc}) we find the entanglement entropy
\be\label{smaster}
S_A = \frac{c}{6}\log\left[ (\ell_1-\ell_2)^2 \frac{(z \bz)^{ \frac{1}{2}(1-\alpha_\psi)}(1-z^{\alpha_\psi})(1-\bz^{\alpha_\psi})}{\alpha_\psi^2(1-z)(1-\bz)} \right]
 \ .
\ee
To fully define this expression we must choose branch cuts, \ie we have the freedom to take $z \to e^{2\pi i n} z$ and $\bz \to e^{2\pi i m }\bz$. The correct choice depends on whether $\psi$ is outside or inside region $A$, so we consider these two options in turn.

\subsubsection{$\psi$ outside region $A$ ($0<\ell_1 < \ell_2$)}\label{ss:outsideq}

At early times $0 < t < \ell_1$, the cross ratio approaches $1$. In order for the correlation function to have the usual Euclidean singularity at $z\sim 1$, $\bz \sim 1$ we must choose the standard branch cut for the powers in (\ref{smaster}).  For small $\delta$ the resulting entanglement entropy is
\be\label{saoutv}
S_A = S_A^{vac} \equiv \frac{c}{3}\log\left(\frac{\ell_2-\ell_1}{\epsilon_{\text UV}}\right) \qquad (0<t<\ell_1) \ .
\ee
The constant depending on the UV cutoff $\epsilon_{\text UV}$ has been fixed to produce the usual vacuum entanglement at $t=0$.

In the window $\ell_1 < t < \ell_2$, the cross-ratios are again near 1. However, from (\ref{zcs}), we see that at $t=\ell_1$, $\bz$ crosses through infinity and its imaginary part changes sign.  Thus $\bz$ moves to another sheet, so we should take $\bz \to e^{2\pi i}\bz$ and then use the standard branch cut in (\ref{smaster}). This gives
\be\label{swind}
S_A = \frac{c}{6}\log\left[\frac{(\ell_2-\ell_1)(t-\ell_1)(\ell_2-t)}{\epsilon_{\text UV}^2 \delta }\frac{\sin (\pi \alpha_\psi)}{\alpha_\psi} \right] \qquad (0< \ell_1 < t <\ell_2 )\ .
\ee
At late times, $t>\ell_2$, $\bz$ crosses back through the branch cut and the answer is once again (\ref{saoutv}).

The results (\ref{saoutv}) and (\ref{swind}) agree with the holographic entanglement entropy on a conical defect geometry computed in \cite{Nozaki:2013wia} (see also \cite{Asplund:2013zba} for explicit expressions in the $\delta \to 0$ limit). While here we have explicitly performed computations at leading order in a small $\delta$ expansion, the agreement with the holographic results of \cite{Nozaki:2013wia} also holds at finite $\delta$. Thus we have derived these results from CFT under our assumption that the identity block dominates the correlator in sparse CFTs.

The discussion of the correct branch cut to reach (\ref{swind}) was brief, so we will now explain how to do this more carefully with the same result.  In Euclidean signature, the 4-point function of local operators does not depend on the operator ordering. This is reflected in the fact that it is a single-valued function of the cross-ratio $z$. In Lorentzian signature, there are branch points when two operators are separated by a null ray, which in our case occurs for $|t| = \ell_1,\ell_2$ (as $\delta \to 0$).  The choice of analytic continuation around these branch points translates into a choice of operator ordering.  For example, consider a Euclidean 2pt function,
\be
D_E(z) = \langle O(z) O(0) \rangle  = \langle O(0) O(z) \rangle \ .
\ee
This is an analytic function of $z$, but setting $z = \sigma + i \tau$, it has a branch cut in the complex time plane starting at $\tau = \pm i \sigma$; which side of the cut we take in the analytic continuation selects the time-ordered or anti-time-ordered Lorentzian correlator. Setting $t \to t - i\epsilon$ is a simple way to select the analytic continuation appropriate for a time-ordered correlator.

\begin{figure}
\centering
\begin{overpic}[scale=1]{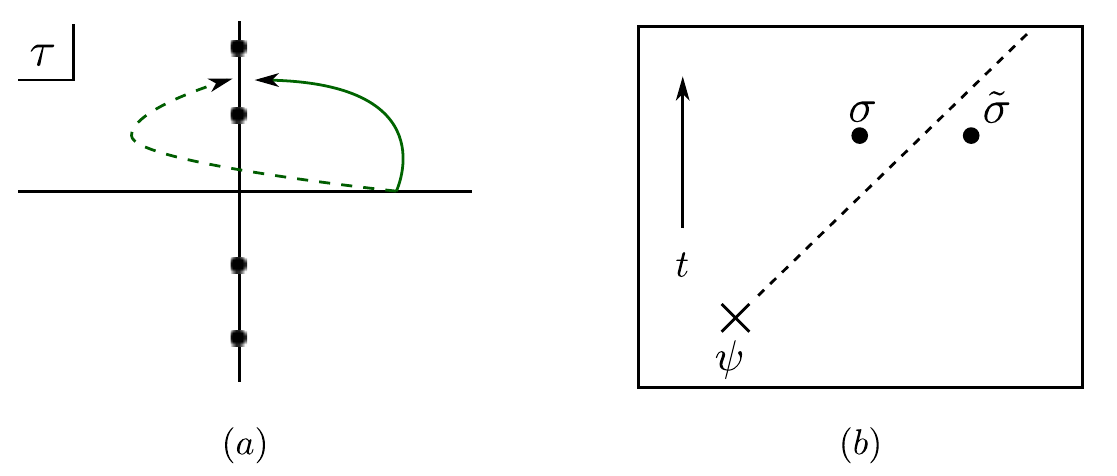}
\end{overpic}
\caption{\textit{Left:} Analytic continuation in the complex time plane. Continuing along the solid curve gives the correlator with the first two operators in time ordering, and the second gives these operators in anti-time-ordering. \textit{Right:} Lorentzian configuration of operators where we cannot use the $\sigma \tsigma$ OPE.\label{fig:cont}}
\end{figure}

In a four-point function, there are multiple branch points corresponding to the various ways of ordering the four operators.  For the Renyi entropy, since we are interested in the expectation value of $\sigma \tsigma$ in the state $|\Psi\rangle$ we must calculate the correlator ordered as
\be\label{ordercor}
\langle \Psi \sigma \tsigma \Psi \rangle \ .
\ee
At early times, all points are spacelike separated. In this case, (\ref{ordercor}) is time ordered, so we can compute it by $t \to t - i \epsilon$ and it agrees with the branch cut used to derive (\ref{saoutv}).  Now suppose $\ell_1 < t < \ell_2$.  Then the time-ordered correlator corresponds to analytically continuing along the solid curve in fig.~\ref{fig:cont}a.
This is related to the expectation value we need by
\be
\langle T \Psi \sigma \tsigma \Psi\rangle = \langle \Psi \sigma \tsigma \Psi \rangle + \langle [\sigma, \Psi] \tsigma \Psi \rangle \ .
\ee
The extra commutator indicates that we should analytically continue along a contour that circles the branch point produced by $\Psi$. This contour is shown as a dashed curve in fig.~\ref{fig:cont}a.  In terms of the cross-ratio, we reach this contour by taking $\bz \to e^{2\pi i }\bz$ with $z$ held fixed.  This was exactly the prescription used to derive (\ref{swind}). (Note that the branch cuts discussed here are different from the branch cuts discussed in section \ref{ss:blockgeo}, see the discussion in section \ref{sss:inside} below.)

A related point is that in Lorentzian signature, it is incorrect to apply the OPE to $\sigma(x_1)\tsigma(x_2)$ `across' the light cone of a third operator, as in fig.~\ref{fig:cont}b.  For $\ell_1 < t< \ell_2$, this means we cannot expand directly in the $\sigma \to \tsigma$ OPE channel, even though $z, \bz \to 1$.  What we have done instead is to expand in the early-time, Euclidean OPE channel $z, \bz \to 1$, then analytically continue to compute the Lorentzian correlator.  Indeed, while the correlator has the usual singularity $(1-\bz)^{-2nH_n}$ for $\bz \to 1$ at early times, it has different behavior as $\bz \to 1$ on the second sheet as we can see from (\ref{smaster}).

In summary, the rise in the entanglement entropy for $\ell_1 < t < \ell_2$ after a local quench is determined by the Virasoro identity block if we first take $\bz \to e^{2\pi i}\bz$ and then expand near $\bz \sim 1$. This is  a braiding operation where we move the operators at $0$ and $\bz$ around each other.\footnote{In \cite{He:2014mwa} the rise in entanglement after a local operator quench in rational CFT was shown to be fixed by the quantum dimension of the operator.  Quantum dimensions are also closely related to braiding of conformal blocks, but braiding appears to be qualitatively different in rational vs.~non-rational CFTs, as we can see by comparing the entanglement entropies. For a different but probably related appearance of the quantum dimension in 3d gravity, see \cite{McGough:2013gka}.} In the limit $n \to 1$ relevant for the entanglement entropy we extracted the behavior after braiding from the explicit expression for the block (\ref{fpsi}). More generally, we do not know a practical way to compute the $n\geq 2$ Renyi entropies at intermediate times, because the needed configuration ($\bz \to 1$ on the second sheet) is very far from the point around which the conformal block is usually expanded ($\bz \to 1$ on the first sheet).\footnote{As written, (\ref{bigfe}) converges only in the unit circle so it cannot be used to compute braiding. However Zamolodchikov's $q$-expansion \cite{Zamolodchikov:1987} (reviewed in \cite{tom}) can be used to evaluate the block on the second sheet, so this could perhaps be used to calculate the Renyi entropy numerically by going to high enough order.}

\subsubsection{$\psi$ inside region $A$ ($\ell_1 < 0 < \ell_2$)}\label{sss:inside}

Only one of the two geodesics drawn in fig.~\ref{fig:defectwinding} played a role in the discussion above, for the quench occurring outside of region $A$.  In CFT language, this is the statement that we only considered a single OPE channel. A new ingredient for $\psi$ inside region $A$ is that there is a second, competing OPE channel which dominates at early times. This is completely analogous to what we found in the static examples of sec.~\ref{ss:blockgeo} and \ref{s:staticex}. 

It is simplest to start at late times and work backwards.  For $t > \ell_2$, both $z$ and $\bz$ are on the principal sheet, so the entanglement entropy is (\ref{saoutv}).  For $|\ell_1| < t < \ell_2$, $z$ crosses the branch cut and the answer is once again (\ref{swind}).

For $t < |\ell_1|$, both $z$ and $\bz$ are on the second sheet.  This would naively lead to the entanglement entropy
\be
S_A = \frac{c}{6} \log\left[ \frac{(\ell_2^2-t^2)(\ell_1^2-t^2)}{ \epsilon_{\text UV}^2 \delta^2}\frac{\sin^2 (\pi \alpha_\psi)}{\alpha_\psi^2}\right] \ ,
\ee
but this answer cannot be entirely correct since it does not reproduce the expected vacuum result for $t \sim 0$.  The issue is that the OPE channel in which we are expanding does not dominate at early times.  When we approximate the full Euclidean correlator $G_n(z,\bz)$ by the identity block in (\ref{snfq}) we must choose a way of analytically continuing around the singularity at $z = 0$, see fig.~\ref{fig:defectwinding} and the discussion in section \ref{ss:blockgeo}.  This is a choice of OPE channel.

Note that this choice of analytic continuation is an issue completely separate from the choices made in the analytic continuation to Lorentzian signature discussed in section \ref{ss:outsideq}. There, the correct analytic continuation was dictated by the operator ordering of the Lorentzian correlator.  In the present case we have two different approximations to the same function $G_n(z,\bz)$ coming from two different OPE channels, already in Euclidean signature, and the correct choice is whichever is larger.

To compute the entanglement entropy in the other channel we take $z \to e^{-2\pi i }z, \bz \to e^{2\pi i}\bz$, and find
\be
S_A = S_A^{vac}
\ee
for $t<|\ell_1|$,
\be
S_A = \frac{c}{6}\log\left[ \frac{(\ell_2-\ell_1)(\ell_2+t)(\ell_1+t)}{\epsilon_{\text UV}^2\delta}\frac{\sin(\pi \alpha_\psi)}{\alpha_\psi}\right] 
\ee
for $|\ell_1|<t<\ell_2$, and
\be
S_A =  \frac{c}{6} \log\left[ \frac{(t^2-\ell_2^2)(\ell_1^2-t^2)}{\epsilon_{\text UV}^2 \delta^2}\frac{\sin^2 (\pi \alpha_\psi)}{\alpha_\psi^2}\right] 
\ee
for $t > \ell_2$.  

Recall that our conjecture is that in a sparse CFT, the correlator is computed by the identity block in the dominant channel. Choosing the dominant channel at each time, the entanglement entropy is therefore
\be
\label{eq:EE_dyn}
S_A = \begin{cases} \vspace{3mm}
\displaystyle\frac{c}{3}\log\left( \frac{\ell_2-\ell_1}{\epsilon_{\text UV}}\right)  &\mbox{\ for \ } \quad t<|\ell_1|, \ t>\ell_2\\  \vspace{3mm}
\displaystyle \frac{c}{6}\log\left[ \frac{(\ell_2-\ell_1)(\ell_2+t)(\ell_1+t)}{\epsilon_{\text UV}^2\delta}\frac{\sin(\pi \alpha_\psi)}{\alpha_\psi}\right] 
 &\mbox{\ for \ } \quad |\ell_1| < t < \sqrt{-\ell_1\ell_2} \\ 
\displaystyle \frac{c}{6}\log\left[\frac{(\ell_2-\ell_1)(t-\ell_1)(\ell_2-t)}{\epsilon_{\text UV}^2 \delta }\frac{\sin (\pi \alpha_\psi)}{\alpha_\psi} \right]  &\mbox{\ for \ } \quad \sqrt{-\ell_1\ell_2}<t<\ell_2
\end{cases} \ .
\ee
This again exactly matches the holographic result of \cite{Nozaki:2013wia} (see also expressions (108)-(111) in \cite{Asplund:2013zba}), with the two OPE channels corresponding to bulk geodesics with different winding around the defect.

\subsection{Joining quench}\label{ss:cc}

In \cite{Calabrese:2007,Calabrese:2009qy} (see also  \cite{Eisler:2007})  Calabrese and Cardy studied a different type of local quench in which two 1+1-dimensional BCFTs are suddenly joined at their boundaries, and subsequently evolve as a single connected CFT. We refer to this specific process as a ``joining quench".

The joining quench is not a local operator quench of the type discussed above, but it shares certain features.\footnote{Note that a local operator quench produces a localized excitation, but is not causal, meaning that information about the quench is not confined inside the lightcone $t>|x|$.  Consider for example the one-point function $\langle\psi |\psi|\psi\rangle$, which differs from the vacuum even at $t=0$. Therefore the local operator quench cannot be produced by a local physical operation at $x=t=0$. This differs from the joining quench, which can be viewed as a sudden local change in the Hamiltonian, and is therefore causal.}  The stress tensor after a joining quench was computed in \cite{Asplund:2013zba} with the result
\be
\label{eq:EM_ev}
	T (w) = -\frac{c}{8} \frac{\delta^2}{(w^2 + \delta^2)^2} .
\ee
This can be compared to the stress tensor created by a heavy operator insertion, (\ref{tprim}) after transforming it with \eqref{conftr}.
We see that the joining quench produces the same stress-energy as a local primary operator of conformal dimension 
\be\label{hjoin}
h = \frac{c}{32} \ .
\ee

Now we can compare the single interval entanglement entropy results of 
\cite{Calabrese:2007, Calabrese:2009qy} to those found 
here, setting $h_\psi = h_{\bar{\psi}} = c/32$, $\alpha_\psi = \tfrac{1}{2}$.  
The results of \cite{Calabrese:2007, Calabrese:2009qy} are universal (independent of the CFT) when the identity operator dominates the OPE.  This is the case for $t< |\ell_1|, t > \ell_2$ if $\ell_2 \gg |\ell_1|$, and for any time if $\ell_1 \gg \ell_2 -\ell_1 > 0$. In these regimes, the results agree with the expressions we find above, except at early times ($t < |\ell_1|$) when $\ell_2 \gg |\ell_1|$. 
In this case, the result of \cite{Calabrese:2007, Calabrese:2009qy} corresponds to the sum of ground state entanglement entropies for two slits $(0, |\ell_1|)$, $(0, \ell_2)$ in a half line for $\ell_1<0$, or to the ground state entanglement entropy for a slit $(\ell_1, \ell_2)$ in a half line for $\ell_1 >0$. Our computation for $t<|\ell_1|$ gives instead the vacuum entanglement entropy of an interval $(\ell_1, \ell_2)$ in a line. This is not a contradiction as it is a different type of quench; at a technical level, the difference comes from an additional OPE channel that exists in the joining quench due to the fact that it is a BCFT calculation, and this channel dominates at early times.

\end{spacing}
\vspace{1cm}
\textbf{Acknowledgments} We thank Steven Avery, Jan de Boer, Frederik Denef, Veronica Hubeny, Nabil Iqbal, Ying-Hsuan Lin, Don Marolf, Joe Polchinski, Mukund Rangamani, James Sully, Tadashi Takayanagi, and Aron Wall for useful discussions. CA is supported in part by a grant from the John Templeton Foundation and in part by the United States Department of Energy under DOE grant DE-FG02-92-ER40699. The opinions expressed in this publication are those of the authors and do not necessarily reflect the views of the John Templeton Foundation. The work of AB and FG was supported in part by the Belgian Federal Science Policy Office through the Interuniversity Attraction Pole P7/37, by FWO-Vlaanderen through project G.0651.11, by the COST Action MP1210 The String Theory Universe and by the European Science Foundation Holograv Network. AB and FG are FWO-Vlaanderen postdocs. The work of TH was supported by the National Science Foundation under Grants No.~NSF PHY11-25915 and PHYS-1066293, the Kavli Institute for Theoretical Physics, and Cornell University. TH also thanks the Solvay Institute at ULB and the Aspen Center for Physics for hospitality during a portion of this work. 

\appendix 

\section{Geodesics length in the defect geometry}\label{aa:geod}

In this appendix we discuss the computation of geodesic lengths in the singular 3d geometry \eqref{defge} using its local equivalence  to AdS$_3$.

In 3d all Euclidean solutions to the vacuum Einstein equations with negative cosmological constant  (away from possible singularities) can be written as \cite{Banados:1998gg}
\be \label{E3d}
ds^2 = \frac{L}{2}dz^2 + \frac{\bL}{2}d\bar z^2 + \left(\frac{1}{y^2} + \frac{y^2}{4}L\bL\right) dz d\bz +  \frac{dy^2}{y^2} \ , 
\ee
with $L=L(z)$, $\bL = \bL(\bz)$. $L$ is related to the dual CFT stress tensor $T$ by
\be
T =  -\frac{c}{12 } L\ .
\ee
The CFT stress tensor transforms under conformal mappings as 
\be \label{Ttransf}
T(z)=  \(\frac{dw}{dz}\)^2 T(w(z)) + \frac{c}{12} \{w,z \}\ ,
\ee
where
\be
\{w,z \}  = \frac{ w'''(z)w'(z)- \frac{3}{2} w''(z)^2 }{w'(z)^2 }\, .
\ee
Correspondingly, the two geometries dual to such CFT states are related via a diffeomorphism that extends the boundary conformal mapping into the bulk.

The singular Euclidean geometry \eqref{defge} of section \ref{ss:blockgeo}  corresponds to 
\be \label{Tdef}
T = \frac{h_\psi }{z^2} \ , \qquad \bT = \frac{\bh_\psi}{\bz^2} \ , 
\ee
while pure AdS$_3$ in Poincar\'e coordinates has $T=\bT =0$,
\be 
ds^2 =  \frac{dw d\bw+ du^2}{u^2} \ .
\ee
The 3-parameter family of maps relating the two can be found  from \eqref{Ttransf}:  
\be \label{dmap}
w(z) =  \frac{a_1}{  z^{\alpha_\psi} + a_2} + a_3 \, ,
\ee
where  $\alpha_\psi = \sqrt{1- 24  h_\psi/c}$.
The full non-linear bulk diffeomorphism relating Poincar\'e AdS$_3$ to a generic geometry \eqref{E3d} has been worked out in \cite{Roberts:2012aq}. In particular, this allows computing the length of geodesics in \eqref{defge}, directly from that of AdS geodesics. 

The expression for the length of a geodesic $\tilde\gamma$ in AdS$_3$, anchored  near the boundary at generic points
\be
(u, w) = (u_{\infty+}, w_{\infty+}) \ , \qquad (u, w) = (u_{\infty-}, w_{\infty-}),
\ee
is
\be\label{geoads}
\mathcal{L} _{\tilde\gamma AdS_3} = \log \frac{(w_{\infty+} -w_{\infty-} ) (\bw_{\infty+} -\bw_{\infty-} ) }{u_{\infty+}u_{\infty-}}\, .
\ee
Starting from this, the length of a geodesic $\gamma$ in the singular geometry of \eqref{defge}, anchored near the boundary at 
\be
(y, z) = (y_\infty, z_{\infty+}) \ , \qquad (y,z) = (y_\infty,z_{\infty-})\, ,
\ee
is straightforward to calculate using the expressions in \cite{Roberts:2012aq}. 
One just needs to apply the diffeomorphism constructed from \eqref{dmap} to pull back the boundary points of $\gamma$ to $(u_{\infty\pm}, w_{\infty\pm})$. The asymptotic form of the diffeomorphism in \cite{Roberts:2012aq} suffices,
\bea 
w_{\infty\pm} &\approx& w(z_{\infty\pm}) \label{asymd} \, , \\
u_{\infty\pm} &\approx& y_\infty  \sqrt{w'(z_{\infty\pm}) \bw'(\bz_{\infty\pm})} \label{asymd2} \, .
\eea
Plugging
\be
(y, z) = \( \epsilon_{\text UV}, 1\) \ , \qquad (y,z) = \( \epsilon_{\text UV}, z_0\)\, ,
\ee
into  \eqref{asymd}-\eqref{asymd2} and substituting into \eqref{geoads}, one obtains the geodesics length in the defect geometry \eqref{defge}
\be
\mathcal{L}_\gamma = \log \frac{(1- z_0^{\alpha_\psi}) (1- \bz_0^{\alpha_\psi}) }{ \alpha_\psi^2}+ \frac{ 1- \alpha_\psi }{2} \log z_0 + \frac{ 1- \balpha_\psi }{2} \log \bz_0 - 2 \log \epsilon_{\text UV} \, .
\ee
Notice that in fact this formula captures the length of the various geodesics in the singular geometry \eqref{defge} with equal endpoints but differing in their winding around $z=0$. 
For simplicity, let us fix $a_1 =1, a_2 = a_3 =0$ for which the inverse map of \eqref{dmap} takes the form $z(w)= w^{-1/\alpha_\psi}$.  
Consider in AdS$_3$ two different geodesics $\tilde\gamma_1$, $\tilde\gamma_2$ which extend between $w_{\infty+}$ and $w_{\infty-}=w_1, w_2$ respectively. 
For any $w_2  = w_1 e^{ 2 \pi  i k \alpha_\psi }$ with integer $k$ and non-integer $k \alpha_\psi$, $w_2$ and $w_1$ represent two distinct points in Euclidean AdS$_3$  both mapping  to the same point $z_0$. The corresponding geodesics $\tilde\gamma_1$ and $\tilde\gamma_2$ will then map to  geodesics anchored at the same points in the defect geometry but with different winding around $z=0$.
In other words the formula for $\mathcal{L}_{\gamma}$ involves choosing a branch cut for $z_0^{\alpha_\psi}$, and this choice selects a winding number for the geodesic around the defect. 


\end{document}